\documentstyle[prb,twocolumn,aps,epsf]{revtex}
\begin{document}

\twocolumn[
\hsize\textwidth\columnwidth\hsize\csname@twocolumnfalse\endcsname
\draft
\title{Role of confined phonons in thin film superconductivity}
\author{E. H. \ Hwang and S. \ Das Sarma}
\address{Center for Superconductivity Research, \\
Department of Physics, University of Maryland, College Park,
Maryland  20742-4111 }
\author{M. A. Stroscio}
\address{U. S. Army Research Office, P.O. Box 12211 \\
Research Triangle Park, NC 27709-2211}
\date{\today}
\maketitle

\begin{abstract}

We calculate the critical temperature $T_c$ and the superconducting energy
gaps $\Delta_n$ of a thin film superconductor system, 
where $\Delta_n$ is the superconducting
energy gap of the $n$-th subband. 
Since the quantization of both the
electron energy and  phonon spectrum arises due to dimensional confinement
in one direction, the effective
electron-electron interaction mediated by the quantized confined phonons is 
different from that mediated by the bulk phonon, leading to the 
modification of $T_c$ in the thin film system.
We investigate the dependence of $T_c$ and $\Delta_n$ on the film
thickness $d$ with this modified interaction.


\end{abstract}
\vspace{0.5in}
]

\newpage


Superconductivity in thin films has been studied for the last four decades.
The phenomenon of thin film superconductivity  has its own specific 
peculiar features.
In early investigations\cite{blat,singh,kir,strongin}, the effect of the film 
size on the 
superconducting transition temperature, $T_c$, for thin films was
investigated. 
Experiments have shown  a monotonic increase of the critical temperature,
$T_c$, with decreasing film thickness\cite{strongin}. From 
the theoretical point of view\cite{blat,singh,kir}, the shape 
resonances and the strong thickness dependence of $T_c$ 
are the characteristic features of thin film superconductivity.  
The size quantization of the transverse motion of the electron in 
the film leads to 
an increase of $T_c$ with decreasing film thickness, arising essentially from
an enhanced effective BCS pairing interaction.
The resonance effects are manifest each time one of the $n$ 
two-dimensional (2D) subband energy 
levels, $E_n(d)$, of a film with thickness $d$, 
(for `transverse' motion perpendicular to the film faces or 
along the confinement direction), passes through the Fermi surface 
as the thickness $d$ is varied.
In previous calculations\cite{blat,singh,kir}, the phonon modes were assumed 
to be the same as in the
bulk material and  only the one-dimensional quantum confinement effects 
of electrons were considered; i.e., 
the superconductivity in the thin film was considered to be arising from the 
attractive electron-electron interaction mediated by bulk phonons interacting 
with subband-quantized electrons.
However, phonons in thin films (`slab' phonon modes) also have 
specific characteristic features 
by virtue of definite boundary conditions imposed by confinement in 
thin films.
The phonon dispersion in a thin film undergoes substantial
modification compared with the bulk, and a quantization of the phonon 
spectrum occurs \cite{nabity,auld,dutta}.
The quantization of the phonon spectra has many effects on thin film electronic
properties, which have been extensively studied, particularly
in the context of semiconductor quantum well structures. 
It is the purpose of this paper to reconsider the previous 
calculations\cite{blat} of thin film superconductivity taking into account 
confined size-quantized phonons in the thin film. 
We find the resonant shape of the 
superconducting transition temperature, $T_c$, arises from both  the quantum 
electronic confinement and phonon confinement.

Following the BCS theory \cite{bar}
the critical temperature $T_c \sim 1.14 \omega_D \exp(-1/N_0V_0)$ 
for a bulk BCS superconductor, where $\omega_D$
is the phonon Debye frequency, $N_0$ the electronic 
density of states at the Fermi energy, and 
$V_0$ the effective attractive electron-electron interaction 
mediated by phonon exchange. An increase 
of $N_0V_0$ implies an increase in $T_c$.
For a thin film with thickness $d$, the density of states at the
Fermi energy develops quantum structure due to the confined 2D subbands:
$N_{film}=(2\nu-1) m/4\pi d = 
N_{3D}(2\nu + 1)\pi/2dk_F$, 
where $\nu=1,2, ...$ 
is the occupied subband index. Thus, when we use the bulk electron-phonon
coupling constant $J$ unmodified by any phonon size confinement corrections, 
which is identical for 
all subbands, the critical temperature 
for thin films depends on the film thickness, and decreases exponentially
as the thickness  increases. However,
for a fixed bulk electron density (or, $k_F$), as the film thickness 
increases the higher quantized subbands 
are occupied by Cooper pairs, so that the critical temperature
of the film jumps to a higher value  due to the higher $\nu$, 
arising entirely from the jumps in the density of states as the 
effective Fermi level moves through higher values of the subband index $\nu$.
This implies that with increasing thickness
of the film  the critical temperature of the
film exhibits resonance features. 
In real thin films, therefore, the electron-phonon coupling
constant is different from that in the bulk material due to the quantization
of the phonon dispersion \cite{hwang}, and therefore the simple resonance 
scenario (discussed above) must be modified.
When we consider the confined phonons in the thin film, the electron-phonon
coupling constant is  different for different subbands, and 
the effective coupling strength decreases with increasing subband
index; that is, higher subbands have progressively weaker 
electron-phonon coupling constants. This 
change of the coupling constant gives rise to features in 
the superconducting energy gap of thin films, which have not 
earlier been considered in the literature. 
In this paper we calculate the critical temperature, $T_c$, and superconducting
energy gaps, $\Delta_n$, with the modified electron-electron interaction  
mediated by the confined slab phonons.
The results for the calculated critical temperature obviously depend on 
the slab phonon dispersion, which in turn
depends on the boundary conditions used in the phonon confinement model.
We have followed here the simple approach of Thompson and Blatt \cite{blat}, 
where the boundary has been treated by an infinite wall,
and we use  wavefunctions which vanish at the boundary.
Although this is a highly simplified phonon confinement model it has 
the virtue of being analytically tractable --- one could 
systematically improve upon this model using our theory as the 
starting point. We expect our results to be well-valid qualitatively.

We assume that our superconducting film with a finite width, $d$, is confined
in the $z$ 
direction by an infinite square-well
confinement potential applied at $z=0$ and $z= d$.
We choose the same infinite (one dimensional) square-well confinement for both 
electrons and phonons, as would be appropriate for a free-standing 
thin film.
Using periodic boundary conditions in the $x$ and $y$ directions with 
periodicity distance $L$, we have 
the one-electron wave function and spectrum
\begin{eqnarray}
\phi_{{\bf k},n}({\bf r},z) &=& u_n(z)\exp(i{\bf k \cdot r})/L, \nonumber \\
\epsilon_n({\bf k}) &=& \frac{\hbar^2 k^2}{2m} + E_n.
\end{eqnarray}
Here, ${\bf k}=(k_x, k_y)$, ${\bf r}=(x,y)$ are the 2D wave vector and
position vector in the plane of free motion, and
by solving Schr\"{o}dinger equation for $z$-direction with confinement 
potential we have
\begin{eqnarray}
u_n(z)& =& (2/d)^{1/2}\sin(k_n z), \nonumber \\
E_n & = &\frac{\hbar^2 (k_n)^2}{2m}, 
\end{eqnarray}
where $k_n = n\pi/d$ with $n=$ integer. Thus,
confinement in the $z$ direction leads
to the quantization of electron energy levels into different subbands.
In addition to the quantization of electron energy, 
we take into account the modification in 
the thin film phonon dispersion arising from the
quantization of the phonon spectrum 
\cite{auld,hwang}. The quantization of the phonon spectra leads to 
the change of the conventional electron-phonon interaction.
The specific expression for the electron-phonon coupling in the thin film 
can be obtained on the basis of the general deformation potential electron-phonon interaction theory \cite{grim}
\begin{equation}
H_{\rm ep}=D \int d^3 r  \Psi^{\dagger} (r) \nabla w(r) \Psi (r),
\end{equation}
where 
\begin{equation}
\Psi(r) = \frac{1}{\sqrt{A}}\sum_{k,n}u_n(z)e^{ik\cdot r} c_{kn},
\end{equation}
$A$ is the 2D area, $c_{k,n}$ is the destruction operator for an electron with
momentum $k$ in the n-th subband, $w({\bf r})$ is the lattice displacement
vector, $D$ is the deformation potential coupling constant (which we 
uncritically assume to be the bulk value since not much experimental 
information is available).
The displacement vector $w({\bf r})$ can be obtained by solving 
equations of motions of the elastic continuum with the appropriate boundary 
conditions \cite{auld,hwang}. We skip the details of this elastic 
continuum theory (which involves substantial algebra) for the sake of brevity

Within the framework of the BCS theory, Cooper pairs are produced consisting 
of electrons with opposite {\bf k} and 
identical n in a thin superconducting film, i.e.,
the electron pairs are produced between 
$({\bf k},n, \sigma)$ and $(-{\bf k},n, -\sigma)$.
The direct effective electron-electron
interaction due to the exchange of virtual confined slab phonons becomes
\begin{equation}
V_{nm}(q) = \sum_{l=1}^{l_{max}}M_{l}^{nm}(q)D_l(q)M_l^{mn},
\label{vnmq}
\end{equation}
where $M_{l}^{nm}$ is the electron-confined phonon matrix element and $D_{l}$
the phonon propagator \cite{hwang}, and the $l$ sums cover all phonons with 
energy less than the cutoff energy, $\omega_D$. 
The maximum value of $l$ contributing to the sum over the slab phonon mode 
$l$ in Eq. (\ref{vnmq}), $l_{max}$, is
given by the condition $l_{max} = ({d}/{\pi})({\hbar \omega_D}/{c})$,
where $c$ is the velocity of the phonons. (We assume that all slab phonons 
have the same velocity because information on the slab mode velocity is not experimentally available.) We then get, 
\begin{equation}
V_{nm} =-\frac{J}{Ad}\sum_{l=1}^{l_{max}} \left [ \beta_{n,n'}^{(l)} 
\right ]^2,
\end{equation}
where the confinement form factor is given by
\begin{eqnarray}
\beta_{n,n'}^{(l)}& = \frac{2}{d}&\int_0^ddz u^*_n(z) \sin(l\pi z/d) u_{n'}(z) 
\nonumber \\
                  & = &\frac{2}{\pi}\left [\frac{l}{l^2-(n-n')^2} - \frac{l}
{l^2-(n+n')^2} \right ].
\label{beta}
\end{eqnarray}
Eq. (\ref{beta}) has the following selection rule: for $|n\pm n'|$ = even (odd)
only odd (even) $l$'s are allowed in the sum over slab modes $l$.
The slab phonons with odd $l$ are symmetric and the ones with even $l$ are
antisymmetric with respect to reflection through $z=d/2$, i.e., under the
transformation $|z-d/2| \rightarrow |d/2 -z|$.
Since the electron wave functions are either symmetric or antisymmetric, 
the couplings between two subbands of the same symmetry 
involve symmetric quantized phonons, 
while couplings between two subbands of different symmetry 
involve antisymmetric phonons. 
(Note that this simplicity will be lost if parity is not a good quantum 
number as it is in our simple infinite square well model, but would not 
be under an asymmetric confinement.) We see that 
$V_{nm}$ decreases with increasing $l$, since the transition of electrons 
to higher subbands cannot be induced by phonons with small momenta $q$.
Thus, the components $V_{nm}$ of the interaction matrix form a 
monotonically decreasing sequence with increasing subband index.
When all the confined phonons contribute to $V_{nm}$ 
(i.e., $l_{max} \rightarrow
\infty$) we recover the bulk phonon mediated results \cite{blat}, i.e.,
$V_{nm} = -(J/Ad)(1 + \frac{1}{2}\delta_{nn'})$, as we should.
As a result of the confined phonons, a superconducting condensate of 
Cooper pairs can be produced in a given subband, 
first because of the attraction  due to the 
electrons in the same subband and second because of transitions from 
other subbands contributing to the condensate.

In the presence
of a number of subbands the reduced BCS Hamiltonian of the system is given by
\begin{eqnarray}
H &=& \sum_{{\rm k},n,\sigma}\xi_n({\rm k})c_{{\rm k}n\sigma}^{\dagger}
c_{{\rm k}n\sigma} \nonumber \\
&+& \sum_{{\rm k},{\rm k}',\sigma}\sum_{n,m}V_{nm}
c_{{\rm k}'m\sigma}^{\dagger}c_{-{\rm k}'m-\sigma}^{\dagger}c_{{\rm k}n\sigma} 
c_{-{\rm k}n-\sigma},
\end{eqnarray}
where $c_{{\rm k}n\sigma}^{\dagger}$ is electron creation operator in the n-th
subband with spin $\sigma$, $\xi_n({\rm k})= \epsilon_n({\rm k}) - \mu $ 
the electron
energy in the n-th subband measured from the chemical potential $\mu$, and
$V_{nm}$ the attractive interaction between $n$-th subband and $m$-th subband
mediated by the confined phonons in the film.

In the BCS theory (which is what we utilize) the gap function has the 
same energy cutoff $\hbar
\omega_D$ as the interaction.
In the weak-coupling approximation we have the 
the superconducting energy gap for $n$-th subband given by the gap equation
\begin{equation}
\Delta_n(T) = \sum_{m{\bf k}}\frac{V_{nm}}{2V}\frac{\Delta_m \tanh(E_m/2kT)}
{E_m},
\end{equation}
where $E_n = (\xi_n^2 + \Delta_n^2)^{1/2}$.
The nondiagonal terms in the sum reflect the possibility of the transition 
of the 
electron pair from one subband into another as a result of interaction with
confined phonons.
Integration over {\rm k} gives the gap function of the subband $n$ at 
$T=0$ K
\begin{equation}
\Delta_n = \frac{Jm}{2\pi}\sum_{n'}\sinh^{-1}\left (\frac{\omega_c}
{\Delta_{n'}}\right ) \Delta_{n'}\alpha_{nn'},
\label{gap}
\end{equation}
where $\alpha_{nn'} = V_{nn'}/J$.
If all  confined phonons contribute equally to the electron-electron 
interaction, i.e.,
($\alpha_{nm}= 1 + \delta_{nn'}/2$), then we recover the 
results of ref. \onlinecite{blat}.  
With the coupled subband interactions Eq. (\ref{gap}) becomes a
non-linear coupled subband matrix equation.
The critical temperature is given by
\begin{equation}
T_c=1.14\omega_D \exp \left (-\frac{2\pi d}{Jm\sum_{n'}\alpha_{1n'}x_{1n'}} 
\right ),
\label{tc}
\end{equation}
where $x_{1n}=\Delta_n/\Delta_1$, the ratio of the n-th subband energy gap 
to the lowest subband energy gap.
For any given finite width, $d$, of the slab, only a finite number of 
eigenvalues, $\xi_n$, contribute; values of $\xi_n$ in excess of $\mu + 
\hbar \omega_D$ make vanishing contributions, because then all 
the $\varepsilon_{\rm k}$'s 
lie outside the interaction region.
Thus, the summation in Eqs. (\ref{gap}) and (\ref{tc}) is only
over all the occupied subbands. For a fixed electron 
density we can find the maximum value
of the $n$ (or, the highest occupied subband) from the chemical potential.
The number density $n_0=N/V$ is related to the chemical potential by the 
relation, $N = 2 \sum_{{\bf k},n} n_{{\bf k},n}$,
where $n_{{\bf k},n} =[\exp(-\varepsilon_n({\bf k})/k_B T) +1]^{-1} $ is 
the Fermi distribution function. At $T=0 K$ we have 
\begin{equation}
\mu = (\pi d \hbar^2/\nu M)\left \{ n_0 + \frac{\pi}{6d^3} \nu (\nu + 
\frac{1}{2}) (\nu + 1) \right \},
\end{equation} 
where $\nu$ is the maximum value of occupied subband $n$ and is given 
by the integral value of the expression, $\nu = {d k_F}/{\pi}$.
If we write Eq. (\ref{tc}) in the usual BCS form, $T_c = 1.14 \omega_D
\exp(-1/N_{film}J_{eff})$, the effective interaction parameter can be written
as
\begin{equation}
J_{eff} = \frac{2J}{2\nu + 1}\sum_{n=1}^{\nu}\alpha_{1n}x_{1n}.
\end{equation} 
As $d \rightarrow \infty$, $J_{eff} \rightarrow J$ because $\alpha_{1n} (
d\rightarrow \infty)= 1 + \delta_{1n}/2$ and $x_{1n}(d\rightarrow \infty)=1$,
and $N_{film} \rightarrow N_{3D}$. Thus, we recover the three dimensional 
results as we expect.

In Fig. 1 we show the calculated superconducting energy gaps of 
each subband and  the
critical temperature as a function of the film thickness $d$ up to the 4-th 
subband occupation.
In this figure we use  the following parameters: Debye energy 
$\hbar \omega_D = 100 K$, 
electron density $n_0=2 \times 10^{22} cm^{-3}$, and $\rho = N_{3D} J = 
mk_F J/(2\pi^2)=0.3 $. In previous calculations\cite{blat,singh} 
the energy gaps  of different subbands 
were find to be the same by virtue of the bulk phonon approximation.
The shape resonance feature in the earlier calculations \cite{blat,singh}
arise only from the effective 2D density of state of the film as
the chemical potential passes through different subbands.
In our calculation we find that the energy gaps are different for different
subbands since the effective coupling strength depends explicitly on the 
occupied subbands. In addition, the resonance structures in our results 
(the sharp maxima in Fig. 1) arise from  both the thickness dependent density 
of states $N_{film}$ and the effective interaction parameter $J_{eff}$.
The energy gap is maximum for the lowest subband and decreases 
in the higher subbands for a fixed film thickness. In Fig. 1(b) we also
compare our results with Ref. \onlinecite{blat} for equivalent parameter 
values. Our results exhibit more 
resonance features and has much lower $T_c$ (typically, about half around 
the maxima) than that of Ref. \onlinecite{blat}.  
Note that any inhomogeneity on the microscopic scale (quite unavoidable 
in real thin films of 5-15 $\AA$ thickness) will considerably suppress 
the resonance features of Fig. 1, and any enhancement in $\Delta_n$ or 
$T_c$ may remain unobservable unless the films are microscopically of 
uniform thickness.
The reduction of the critical temperature
in our calculation can be explained by the enhancement of the effective
interaction parameter in the slab phonon model. 
The present work can easily be extended to include more 
realistic boundary conditions\cite{singh}, but one then needs to 
resort to numerical work right from the beginning, losing much of 
the essential qualitative physics of the phenomenon. We believe that 
the basic physics discussed in this paper and the qualitative features 
of our results shown in Fig. 1 transcend our specific model, and should 
be valid in any BCS type superconductivity in thin films.

This work is supported by the U.S.-ARO and the U.S.-ONR.

\begin{figure}
\epsfysize=13.1cm
\epsffile{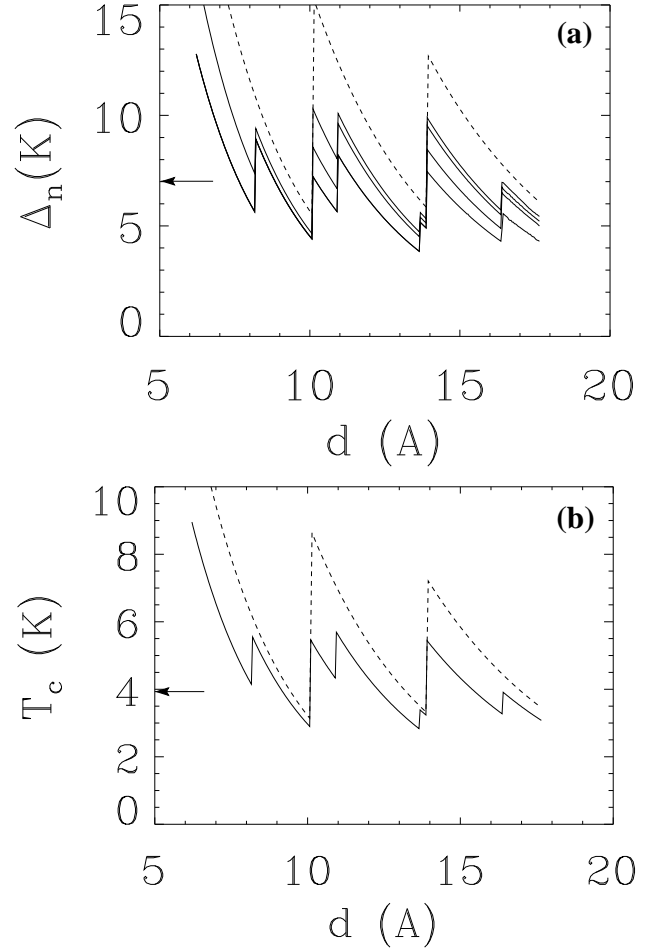}
\vspace{0.5cm}
\caption{
(a) The calculated superconducting energy gaps, 
$\Delta_n$, of each occupied subband $n$, and (b) the 
critical temperature for an electron density $n=2 \times 10^{22} cm^{-3}$
as a function of the film thickness $d$. In (a) the highest curve is the 
energy gap of the lowest subband ($n=1$), and the second highest curve is that
of the first excited subband ($n=2$), and so on, with four occupied subbands. 
In (b) 
the solid line indicates our result with the interaction mediated by 
confined slab phonons and the dashed
line from ref. [1] corresponds to the bulk phonon result. The arrow 
in each figure indicates the purely bulk result.
}

\end{figure}

\end{document}